\documentclass[12pt,preprint]{aastex}
\usepackage{lscape,graphicx}

\def\fluxu{\,{\rm erg} \ {\rm cm}^{-2} \ {\rm s}^{-1}}

\def\lsim{\mathrel{\hbox{\rlap{\lower.55ex \hbox {$\sim$}}\kern-.0em 
\raise.4ex \hbox{$<$}}}}  
\def\gsim{\mathrel{\hbox{\rlap{\lower.55ex \hbox {$\sim$}}\kern-.0em 
\raise.4ex \hbox{$>$}}}}

\def\arcmin{\hbox{$^\prime$}}
\def\arcsec{\hbox{$^{\prime\prime}$}}

\def\hardrange{2 -- 10 keV}
\def\softrange{0.5 -- 2 keV}
\def\hardfluxrange{S_{\rm 2 - 10\, keV}}

\def\lognlogs{$\log N - \log S$}
\def\fxfo{\log(f_{\rm x}/f_{\rm o})}
\def\spose#1{\hbox to 0pt{#1\hss}}
\def\simlt{\mathrel{\spose{\lower 3pt\hbox{$\mathchar"218$}}
     \raise 2.0pt\hbox{$\mathchar"13C$}}}
\def\simgt{\mathrel{\spose{\lower 3pt\hbox{$\mathchar"218$}}
     \raise 2.0pt\hbox{$\mathchar"13E$}}}

\begin{document} 

\title{The Serendipitous Extragalactic X-Ray Source \\ Identification
(SEXSI) Program: II. Optical Imaging}

\author{Megan E. Eckart\altaffilmark{1}, 
Elise S. Laird\altaffilmark{2}$^{,}$\altaffilmark{3},
Daniel Stern\altaffilmark{4}, \\
Peter H. Mao\altaffilmark{1}$^{,}$\altaffilmark{5}, 
David J. Helfand\altaffilmark{2}, \& Fiona A. Harrison\altaffilmark{1}}

\begin{abstract}
The Serendipitous Extragalactic X-ray Source Identification (SEXSI)
Program is designed to expand significantly the sample of identified
extragalactic hard X-ray sources at intermediate fluxes, 
$10^{-13} \fluxu \simlt S_{\rm 2-10~keV} < 10^{-15} \fluxu$.  
SEXSI, which includes
sources derived from more than 2 deg$^2$ of {\em Chandra} images,
provides the largest hard X-ray-selected sample yet studied, offering
an essential complement to the {\em Chandra} Deep Fields (total area
$\sim 0.2$~deg$^2$).  In this paper we describe $R$-band optical
imaging of the SEXSI fields from the Palomar, MDM, and Keck
observatories.  We have identified counterparts, or derived flux limits
for nearly 1000 hard X-ray sources.  Using the optical images, we
derive accurate source positions.  We investigate correlations between
optical and X-ray flux, and optical flux and X-ray hardness ratio.  We
also study the density of optical sources surrounding X-ray
counterparts, as well as the properties of optically-faint, hard X-ray
sources.

\end{abstract}


\altaffiltext{1}{Space Radiation Laboratory, 220-47, California Institute
of Technology, Pasadena, CA 91125} 
\altaffiltext{2}{Columbia University, Department of Astronomy, 550 West
120th Street, New York, NY 10027} 
\altaffiltext{3}{Astrophysics Group, Imperial College London,
Blackett Laboratory, Prince Consort Road, London SW7 2BW, UK}
\altaffiltext{4}{Jet Propulsion Laboratory, California Institute of
Technology, Mail Stop 169-506, Pasadena, CA 91109} 
\altaffiltext{5}{University of California at Los Angeles, Department
of Earth and Space Sciences, 595 Charles Young Dr. East, 
Los Angeles, CA 90095}

\section{Introduction}
\label{sec:intro}

With the successful launch in 1999 of the {\em Chandra X-Ray
Observatory}~\citep{Weisskopf:96} came the opportunity to explore the
X-ray universe with greater angular resolution and sensitivity in the
\hardrange~energy range than ever before. One of the primary goals of
the mission is to perform surveys of the extragalactic sky at these
higher energies.

The longest {\em Chandra} pointings thus far are the {\em Chandra} Deep Field -
North~\citep{Alexander:03} and -South~\citep{Giacconi:02}, with 2~Ms
and 1~Ms exposure times, respectively. These fields have explored faint
sources, with many objects reaching below the luminosity range of
active galactic nuclei (AGN) to include ``normal" galaxies. Though the
Deep Fields are pushing to lower fluxes than previously attainable,
they cover only a small area ($\sim 0.1$ deg$^2$ each), and do not have
good statistics for the flux range in which the dominant contribution
to the \hardrange~X-ray background arises. To get such statistics, a
larger area of the sky must be covered by combining multiple {\em
Chandra} pointings.

The {\em Chandra} archive provides the opportunity to explore multiple,
mid-depth ($\sim 50 - 100$ ks) exposures.  In \citet[][Paper~I]
{Harrison:03} we presented the Serendipitous Extragalactic X-ray Source
Identification (SEXSI) program design, our X-ray analysis techniques and
a catalog of over one thousand \hardrange~sources, as well as some
initial X-ray results and comparisons with previous work.  With data
from 27 archival {\em Chandra} fields, SEXSI is designed to focus on a
large-area ($\sim 2$ deg$^2$) sample of X-ray sources in the
intermediate flux range ($\sim 10^{-13} - 10^{-15} \fluxu$). 
Our goal is to determine spectroscopically the redshift 
distributions and nature of hard (\hardrange) X-ray source populations 
from which the bulk of the X-ray background arises. The campaign
of single-color photometry presented here attempts a minimum
depth for every field of $R \sim 22 - 23$, consistent with identifying 
counterparts for which classifiable spectra can be obtained with a 10-m telescope in 
a $\sim 1$ hour integration.
This strategy maximizes the number of hard sources  identified 
spectroscopically in our fields for our fixed observing campaign.

Other programs are also aiming to understand the X-ray source 
population at medium depth with degree-scale {\em Chandra} and 
{\em XMM-Newton} surveys. Examples of such surveys include HELLAS2XMM
\citep[e.g.,][]{Baldi:02,Brusa:03} and the {\em Chandra} 
Multiwavelength Project \citep[ChaMP; e.g.,][]{Kim:03, Green:04}. 
\citet{Brandt:04} presents a comprehensive list of 
the many ongoing X-ray background surveys.

The SEXSI program is unique among the {\em Chandra} Serendipitous
surveys in concentrating on obtaining a uniform 
spectroscopic survey restricted to the hard source population.   
The ChaMP, in contrast, is a
comprehensive program of imaging and spectroscopy to follow-up both soft
and hard sources in a large number of publicly-accessible
extragalactic {\em Chandra} pointings.   The SEXSI spectroscopic catalog will
therefore contain a larger fraction of obscured AGN than 
ChaMP.  Eight of the fields are common to both surveys,  and for
one of these ChaMP has published some spectroscopic followup \citep{Green:04}.
Even for the common fields, SEXSI plans deeper spectroscopy and 
should identify a greater number of hard X-ray sources.
Between the large number of distinct fields and the
different focus for ground-based followup, the SEXSI survey provides
an essential complement to current surveys of the sources
responsible for the X-ray background.

In this paper, we present optical $R$-band imaging of the SEXSI {\em
Chandra} fields, as well as the methodology used to derive X-ray source
identifications and a catalog of $R$-band counterpart magnitudes. In
addition, we explore relationships between the X-ray flux, hardness
ratio, and $R$-band magnitude.  A third paper in this series (Eckart
et al., in preparation; Paper III) will include our optical
spectroscopic results, with redshifts and classifications for $\sim 450$
of these objects, and a discussion of the luminosity distribution,
redshift distribution, and composition of the sample.

\section{Optical Imaging and Reduction}
\label{sec:images}

We have imaged the SEXSI fields using the Palomar 60-inch (P60) and
200-inch (P200) telescopes, the MDM 2.4-meter (MDM-2.4) and 1.3-meter
(MDM-1.3) telescopes, and the Keck~I telescope.  Much of the early
optical imaging for this program was done with small field-of-view,
single-CCD array cameras.  For example, the P60 CCD13 camera has a
12.6\arcmin$\times$12.6\arcmin\ field of view.  This requires several
pointings to map a typical $17\arcmin \times 17\arcmin$\ {\em
Chandra} field, making multi-color imaging very time consuming.
Since Spring 2000, both the P200 and MDM-2.4 have had large-format
($\simgt 24\arcmin \times 24\arcmin$) CCD cameras available, allowing
us to cover nearly all of a {\em Chandra} field in a single pointing.
The P200 Large Format Camera \citep[LFC;][]{Simcoe:00} uses the Sloan
digital sky survey (SDSS) filter set \citep[$g'$, $r'$, $i'$, and
$z'$;][]{Fukugita:96}, while, for our MDM-2.4 8K camera \citep{Crotts:01}
imaging, we used Johnson-Cousins filters ($B$, $V$, $R$, and $I$).
In each filter band, our goal has been to obtain images with a
limiting Vega magnitude of $\sim 24$, the practical limit for obtaining
counterpart spectra at the Keck telescope in reasonable integration times.
Because the large-format cameras were not available for the first 1.5
years of the SEXSI project, some fields are imaged only in $R$ band
(due to observing time constraints), while the fields observed later
in the program generally have multicolor data.  For the purposes of
this paper, we use only the $R$-band data.  Table~\ref{tbl:cameras}
summarizes the characteristics of the telescopes and imaging instruments
used in this program.

We reduced the optical images using standard techniques, relying upon
IRAF tasks for the preliminary steps.  Median-combined bias frames, taken
the same night as the science data, were subtracted from the images.
For flatfielding, we used either dome flats or sky flats generated by
combining all the images from a night, including twilight flats, with
min/max rejection to remove stars.  We used standard techniques to mask
out cosmic rays and bad pixels prior to combining data.

For astrometric calibration of most of the data, we used the {\tt
DOPHOT} software package to locate all nonsaturated, point-like objects
in individual exposures.  Comparing with the USNO-A2.0 catalog
\citep{Monet:98}, we then used the object-matching program {\tt
starmatch} (by Doug Reynolds) to both align the images before stacking and
to obtain an astrometric solution for the final, combined image.  The
standard deviations of the astrometric solutions are generally $\le
0.3\arcsec$ in each axis.

We reduced images using the same methodology for both
the LFC data from the P200 telescope and the 8K data from the MDM-2.4
telescope, although slightly different suites of software were 
required.
We relied upon the {\tt MSCRED} package within IRAF, designed for
reduction of large-format, multi-array optical data from the Kitt
Peak/Cerra Tololo MOSAIC cameras, following the informative reduction
pipeline outlined by B.~Jannuzi\footnote{See {\tt
http://www.noao.edu/noao/noaodeep/Reduction0pt/frames.html}.}.  For the
LFC images this
necessitated first changing the data into the requisite format using
the {\tt LFCRED} package created by M.~Hunt\footnote{See {\tt
http://wopr.caltech.edu/$\sim$mph/lfcred}.}.

\subsection{Photometric calibration}

We relied on a variety of methods to provide photometric calibration of
the data, depending upon the conditions under which data for a SEXSI
field were taken.  Throughout, unless otherwise noted, we refer our
magnitudes to the Vega-based system.  If conditions were photometric, we
determined magnitude zeropoints from \citet{Landolt:92} fields observed
the same night, at an airmass close to that of the field in question.
In cases where a non-photometric image overlaps a photometric image
from another night, we determined the zeropoint magnitude by matching
the photometry of overlapping regions.

In cases where we obtained no photometric images of a field, we
relied on shallow sky surveys to provide magnitude zeropoints.  Where
available, we used $r'$ photometry from the SDSS first data release
\citep{Abazajian:03}, converting from AB magnitudes to Vega magnitudes
with 
\begin{equation}
r'({\rm Vega}) = r'({\rm AB}) - 0.17. 
\end{equation}
Alternatively, we used the USNO-B1.0 catalog \citep{Monet:03}
to establish the zeropoint.  Typical photometric calibration errors for
this technique were 0.2 mag.

For the purposes of this paper, we make the assumption that the SDSS
$r'$ and Johnson-Cousins $R$ filters are similar.  Convolving the
respective filter transmission curves with the \citet{VandenBerk:01}
composite SDSS quasar spectrum redshifted from $z = 0$ to $z = 3$, we
find that this assumption provides a $\simlt 0.1$ mag systematic
uncertainty in our photometry.  Performing the same exercise on the
\citet{Kinney:96} composite galaxy templates, we find systematic
offsets $\simlt 0.2$ mag for $z = 0$, with larger systematic offsets at
higher redshifts.  For example, the S0 composite template at $z = 1$
has a 0.39~mag difference between the $r'$ and $R$ Vega magnitudes due
to the 4000 \AA\  break redshifting out of the $r'$ filter.  Though
imprecise, our assumption that the $r'$ and $R$ filters are similar is 
adequate to establish the faint source limit of each image, and to plan
spectroscopic observing runs.

\subsection{Source extraction and limiting magnitudes}
\label{sec:sextractor}

For extraction of sources from our combined, calibrated optical images,
we used the SExtractor code of \citet{Bertin:96} to generate catalogs
for both the full optical images and for the smaller, central portions
of each image.  Because they avoid the CCD edges where the increased
noise leads to spurious sources, these latter catalogs are used to
derive photometric depths.  Quoted magnitudes refer to the SExtractor
MAGBEST, which usually reports photometry within elliptical apertures
\citep{Kron:80}, but reverts to a corrected isophotal magnitude if a
source is strongly contaminated by nearby sources.

\begin{figure}
\epsscale{1} 
\plotone{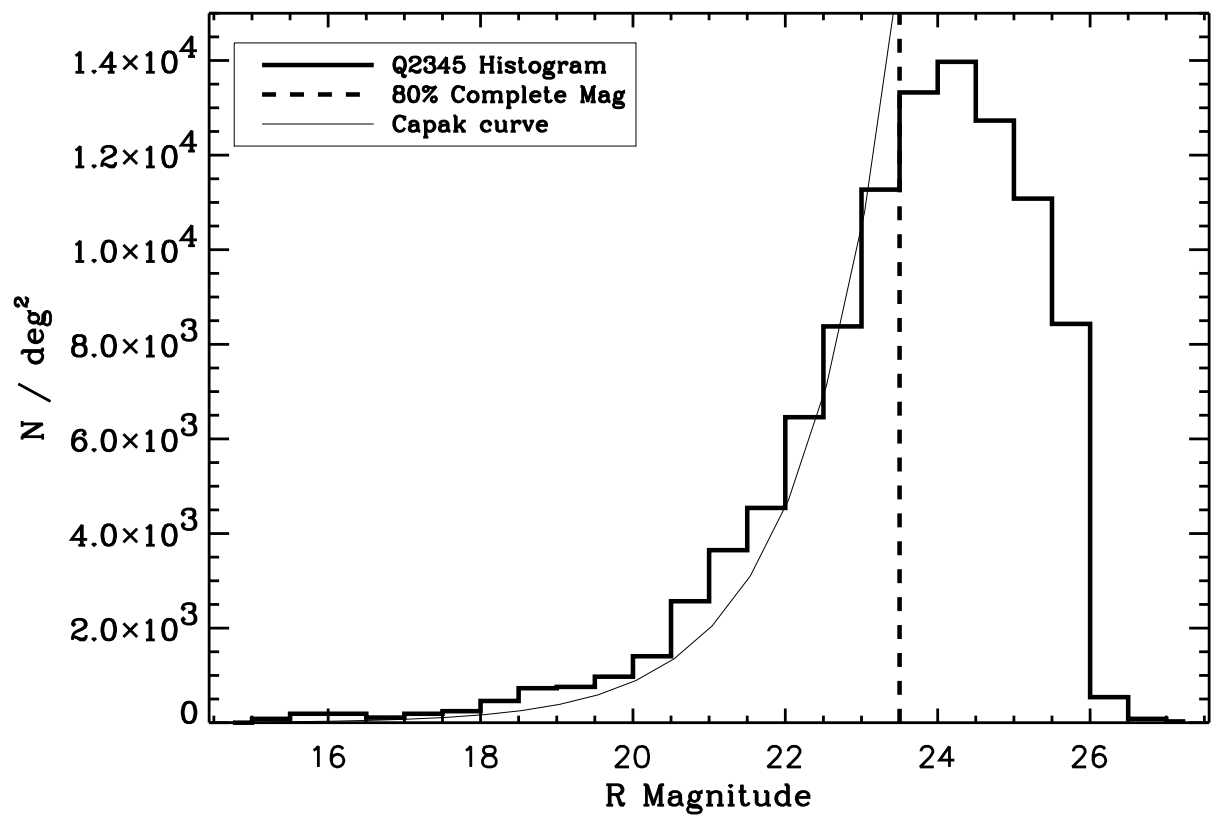} 
\caption{Example of $R$-band number counts for a typical SEXSI image:
the Q2345 number counts from our MDM 2.4-meter/8K image is shown as a
thick, solid histogram.  The~\citet{Capak:04} fit to deep imaging number
counts is shown as a thin, solid line.  Our 80\%\ completeness limit
(vertical dashed line) determination is described in \S~\ref{sec:sextractor}.}

\label{fig:magcutoff}
\end{figure}

To determine limiting magnitudes for each image, we made histograms of the
$R$-band source density in 0.5 magnitude bins.  Figure~\ref{fig:magcutoff}
shows an example histogram from SEXSI field Q2345.  For each SEXSI
field image, we compare the measured number counts $N$ to that derived
by \citet{Capak:04} from deep Subaru imaging of the {\em Hubble} Deep
Field-North:
\begin{equation}
N = B~10^{A*{\rm (AB~Magnitude)}}, 
\end{equation}
\noindent where $N$ has units of number degree$^{-2}$ 0.5 mag$^{-1}$.
For $R$-band magnitudes in the range $20.0 - 25.0$, \citet{Capak:04}
finds $A=0.361$ and $\log{B} = -4.36$.  Converting our $R$-band
photometry into the AB system with Equation~1, we plot our measured
number counts for each image with the published, deep-field fit (see
Figure~\ref{fig:magcutoff}).  In order to derive a crude depth for our
images, we compare our number counts in each 0.5~mag bin to Equation~2
(at bin center).  Limiting magnitudes are defined by the inter-bin flux
where our number counts go from greater than 80\% to less than 80\% of
the \citet{Capak:04} value:  this provides an $\sim 80\%$ completeness depth
for each image.  By the nature of our algorithm, then, these limits
are good to $\approx 0.5$ mag.

Table~\ref{tbl:images} summarizes the $R$-band optical imaging we have
obtained for the SEXSI fields.  The average X-ray to optical offsets,
$\langle \Delta\alpha_{\rm xo} \rangle$ and $\langle \Delta\delta_{\rm
xo} \rangle$, are discussed in \S~\ref{sec:matching} below.  We derived
$R$-band Galactic extinction for the central position of each X-ray
image using the NED Galactic extinction calculator\footnote{See {\tt
http://ned.ipac.caltech.edu/forms/calculator.html}.}.  Checking the
variance of the NED extinction values across the {\em Chandra} field
of view, we find that the typical error on the extinction is $\approx
0.02$ mag.  Three of our fields, however, have significantly higher
extinctions than the other fields, and thus have an uncertainty of $\pm
0.2$ mag in their extinction values for individual sources.  These fields
are flagged in Table~\ref{tbl:images}.

\begin{figure}
\epsscale{1.0} 
\plotone{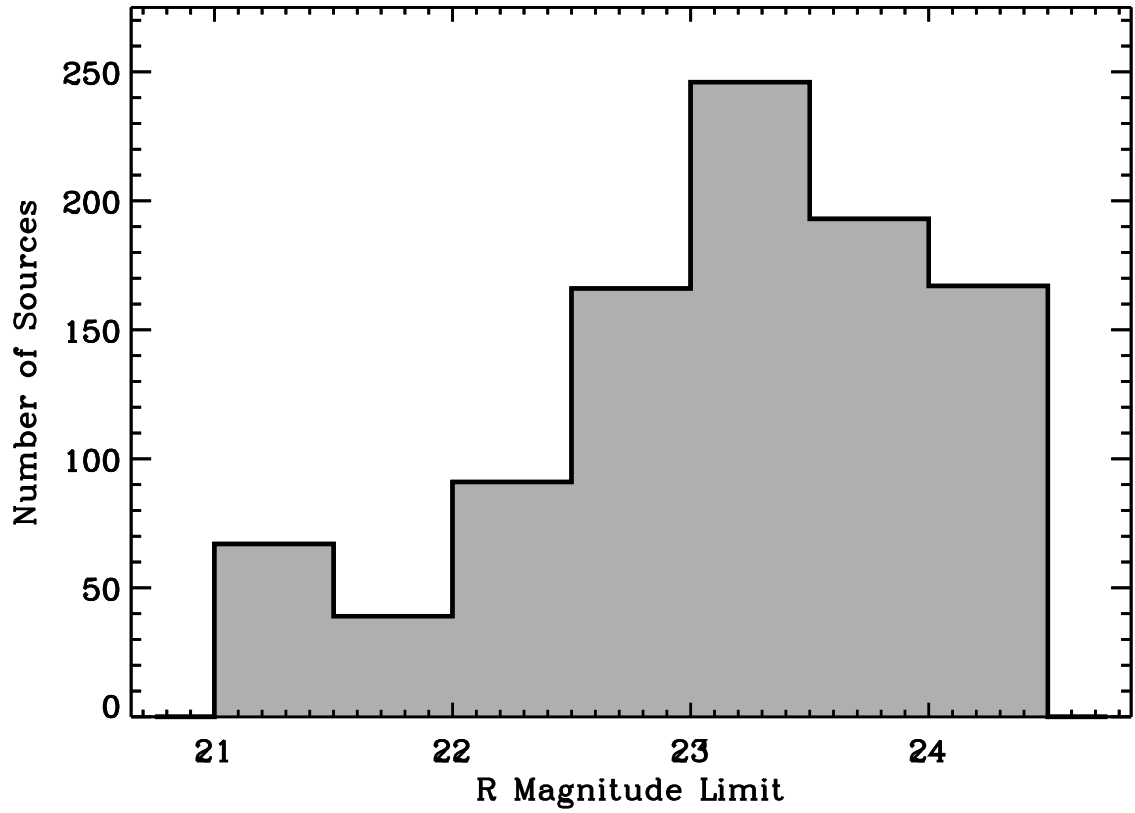} 
\caption{Histogram of limiting optical magnitudes for all SEXSI sources
with optical coverage, excluding sources with non-detections where the 
background is contaminated by a nearby bright source (optflag $=6$, 
see \S~\ref{sec:catalog}). Most sources have limiting magnitudes 
of at least $R=23$.}
\label{fig:depth_hist}
\end{figure}

Approximately half of the SEXSI fields were observed with a single,
uniform depth image; the other half rely on imaging from multiple cameras
and telescopes over several observing runs.  Of the fifteen fields with a
single optical image, six reach $R$-band depths $\geq 24.0$, and ten reach
$R$-band depths $\geq 23.2$.  The shallowest field, CL~0442+0202, only
reaches $R = 21.1$, while the deepest field, CL~0848+4454, reaches $R = 24.4$.
Of the twelve fields with imaging from multiple pointings,
eight have sections reaching depths of at least $R = 23.0$.  A histogram of imaging depth for each source
is shown in Figure~\ref{fig:depth_hist}.

\section{Optical Counterparts to X-Ray Sources}
\label{sec:matching}

We match X-ray and optical sources in several steps. The procedure is
slightly different depending on whether there is one optical image or
multiple optical images per {\em Chandra} field.

For fields with a single optical image, we first identify the closest
optical source within 4\arcsec\ of each {\em Chandra} source position.
Excluding sources displaced by more than 1.5$\sigma$ from the mean offset,
we use these identifications to calculate a first estimate of the average
offset between the {\em Chandra} and USNO astrometry.  For the following
iterations, the search radius used to identify optical counterparts
depends upon the off-axis angle (OAA) of the X-ray source:  we use the
larger of 1.5\arcsec\ and PSF/3, where PSF is the full-width, half-maximum of
the {\em Chandra} point spread function at each OAA.  Again, we exclude
sources displaced by more than 1.5$\sigma$ from the mean as we iteratively
correct the X-ray source positions.  We continue iterating until the 
corrections to the offsets are less than the corresponding standard deviation.
Two to three iterations are typically required, with
the biggest shift applied after the initial matching.  As shown in Table~2,
the typical standard deviation on this astrometric correction is $\simlt$
0.3\arcsec, comparable to the accuracy of the optical astrometry.

For fields with multiple optical images we modify this algorithm slightly.
Some images cover an area with few (or one) X-ray sources, which, if we
were to follow the procedure described above, could lead to incorrect
offsets and matches.  Instead, for a given {\em Chandra} field, we first
take each optical image and find the closest match within a
4\arcsec\ radius of all {\em Chandra} source positions that fall within
the image.  We record these astrometric differences.  For X-ray sources
falling on multiple optical images, we use the data from the image
with the deepest limiting magnitude.  As before, we use this list of
multi-image astrometric differences to calculate an average {\em Chandra}
to USNO offset, again eliminating sources greater than 1.5$\sigma$ from
the mean offset.  This astrometric correction is then applied to all
{\em Chandra} positions for the SEXSI field considered.  A second pass
at optical identifications is then made with the OAA-dependent matching
radius, using the larger of 1.5\arcsec\ and PSF/3. Of the 998 SEXSI
sources with optical coverage, 655 used a 1.5\arcsec\ search
radius, while 343 used the PSF/3 search radius.  

We expect few false matches due to an optical source randomly
overlapping the matching search area. The number of false matches per
field depends upon both the number of X-ray sources detected in the
{\em Chandra} field, and the depth of the optical image. We
tested several fields from Table~\ref{tbl:images} for false matches by
finding the total area covered by the 1.5\arcsec\ radius search circles
and multiplying that area by the optical source density. We find that
for medium-depth X-ray and optical images, the number of false sources
detected per field is generally less than one. For example, we predict
0.6 false matches for Q2345, 0.7 false matches for CL 0442+0202, and
1.3 false matches for 1156+295, where these numbers are for the
1.5\arcsec\ search radius sources only.  For the 34\% of the sources with
larger OAAs and thus larger search areas we expect a larger fraction of
false matches, though still fewer than 3 per field for most fields.
For example, we predict 1.3 false matches from the $>
1.5$\arcsec\ match radius sources in Q2345, 1.4 false matches in CL
0442+0202, and 2.4 false matches in 1156+295.
For several of the fields, including the three example fields just mentioned,
we shifted all X-ray source positions by 1\arcmin, producing a ``fake''
X-ray catalog, and ran the 
matching routine again with the average X-ray-to-optical offset 
forced to zero. This process, repeated with several different
shifts of the X-ray source positions, gives a check on the number 
of false detections predicted above. For all of the fields tested the 
number of false detections are consistent with the values we calculate.
Our spectroscopic results are also consistent 
with this matching scheme and will be discussed further in Paper III.

\begin{figure}
\epsscale{1} 
\plotone{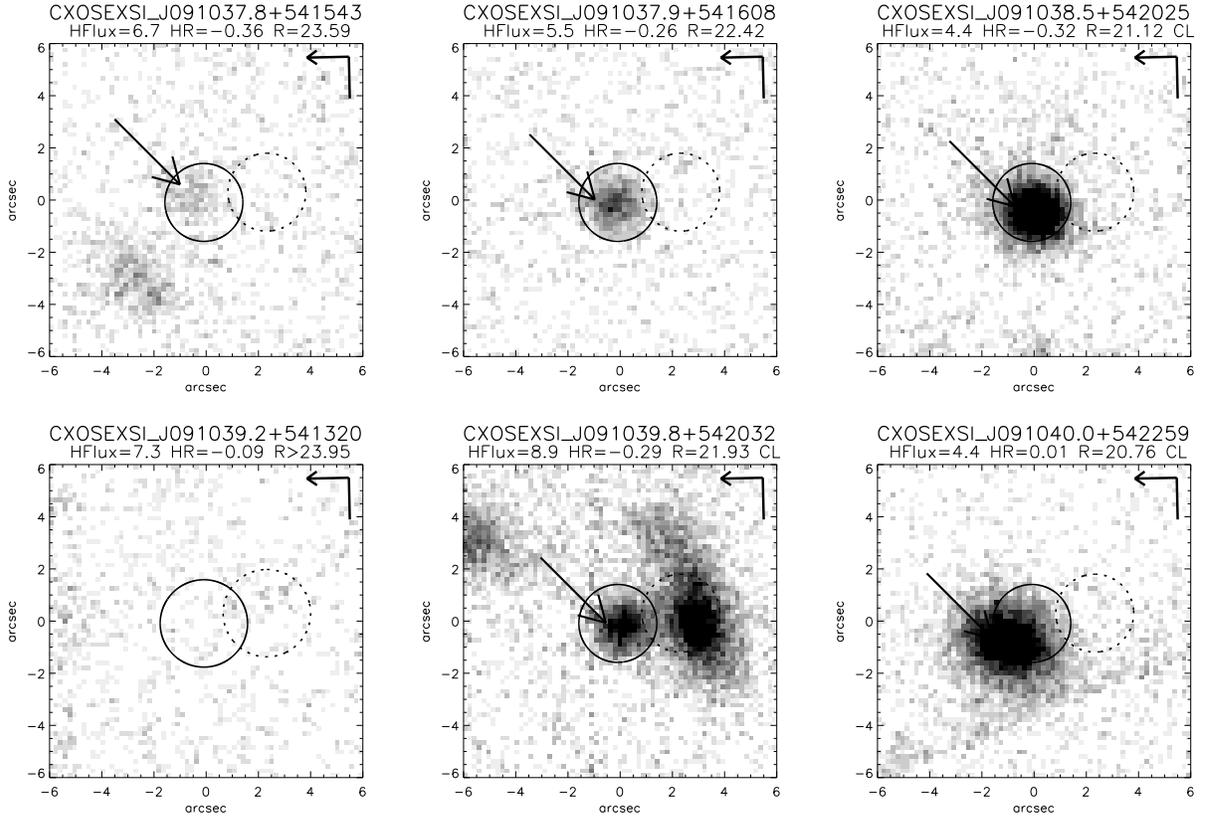} 
\caption{Example of the $R$-band postage stamp cutout images.  Cutouts for
the full catalog of hard-band SEXSI sources with optical follow-up are
presented in the online version of the paper (Figures 3b -- 3do); 
we show representative
images from the RX~J0910 field here.  Images are centered at the
astrometrically-corrected {\em Chandra} source positions, and the
image orientation is shown in the upper-right corners (North has the
arrowhead, East lacks an arrowhead).  The dashed circles are centered on
the original X-ray-derived source positions, while the solid circles show
X-ray source positions corrected for the {\em Chandra} pointing error; circle
radii are a function of X-ray off-axis angle as described in the text.
Arrows point to the optical counterpart if one is present.  Text above
each cutout identifies the source and labels its \hardrange~flux (in
units of $10^{-15} \fluxu$), hardness ratio $HR$ (defined in \S~\ref{sec:catalog}), and
$R$-band magnitude or the limit thereto.  Flags on the $R$-band photometry
are also presented, though no flagged sources are shown in the example figure. 
A dagger ($\dagger$) indicates that the counterpart is near a bright source, affecting 
the $R$-magnitude, a double-dagger ($\ddagger$) flags cases where there is 
more than one optical source within the search area, and an asterisk ($\ast$) 
indicates that the $R$-band magnitude is higher than the limiting magnitude
of the image. Additionally,
CL denotes a source potentially falling within 1~Mpc of a target cluster center
(see \S~\ref{sec:matching}).}

\label{fig:postage}
\end{figure}

Once matching is complete, we produce first-draft optical postage stamp
cutouts for our entire hard X-ray catalog (see
Figure~\ref{fig:postage}).  Each stamp is centered on the
astrometrically-corrected X-ray source position, and the size of the
{\em Chandra} PSF-dependent search radius is shown by the (centered)
solid circle.  A dashed circle, located at the un-corrected {\em
Chandra} position, illustrates the need for this offset correction.
Arrows point to identified optical counterparts, while the absence of
an arrow signals a non-detection.  If the magnitude found by SExtractor
is fainter than the limiting optical image, we annotate the photometry
with an asterisk.
As in Paper~I, a ``CL" flag is added if a source is potentially within
1~Mpc of a target cluster center.

Next we visually inspect postage stamps for each X-ray source, flagging
sources with potentially inaccurate photometry as required.  These flags,
detailed in \S~\ref{sec:catalog}, identify sources which either (1) 
have multiple optical
identifications within the PSF-dependent search area, or (2) are, or
are near, a saturated source in our optical image.  A final matching
iteration is then done, excluding the newly flagged sources.  We produce
new optical catalogs and postage stamp cutouts, omitting stamps for the
handful of saturated sources where counterpart magnitudes are drawn
from the literature and the 65 SEXSI sources that
either lack optical coverage in our data or have an unknown limiting magnitude
due to nearby bright-source contamination (optical flag = 6, 
see \S~\ref{sec:catalog}).  An example of six postage stamp
cutouts is included as Figure~\ref{fig:postage}; the entire catalog of
postage stamps is provided in the online version of the manuscript.

For the 262 sources with $22 < R_{\rm limit} \le 23$,
160 (61\%) have identified counterparts, while for the 434 sources
with $23 < R_{\rm limit} \le 24$, 291 (67\%) have 
identified counterparts, and for the 167 sources with 
$R_{\rm limit} > 24$, 124 (74\%) have identified counterparts.
Our total sample of 947 sources with unambiguous photometry
(no contamination) identifies 603 counterparts (64\%).

\section{The Catalog}
\label{sec:catalog}

In Table~\ref{tbl:catalog} we present the catalog of 1034 hard-band
SEXSI sources -- the table is published in its entirety in the electronic
version of the {\em Astrophysical Journal}.  
Columns~1 -- 7 present X-ray source information for
easy reference, while the optical photometric data are presented in
columns~8 -- 15.  Complete X-ray source information is presented in
Paper~I, Table~4. The X-ray source positions in Table~\ref{tbl:catalog}
are corrected for mean optical to X-ray offsets.  Note that since the
source names (column 1), identical to the source names in Paper~I, are
derived from the hard-band X-ray images, the refined positions of
columns~2 -- 3 will not exactly match those of column~1 (though mean
offsets are typically less than 1\arcsec). Column 4 lists the off-axis
angle (OAA, {\em i.e.}, the angular distance, in arcmin, of the source
position from the telescope aim point).  The \hardrange~flux (in units
of $10^{-15} \fluxu$) and detection SNR are shown in columns~5 -- 6,
while column~7 gives the hardness ratio, $HR = (H-S)/(H+S)$, where $H$
and $S$ are the counts cm$^{-2}$ in the \hardrange~and
\softrange~bands, respectively. Here, as distinct from Paper I, we
record the hardness ratio derived from the net soft X-ray counts recorded at
the hard-band source position when there was not a significant soft-band
source detected (in Paper I these cases are reported as $HR=1.0$). In 
addition, for a subset of these cases, when the soft-band counts
recorded at the hard-band source position
were less than twice the soft-band background counts, 
the $HR$ is considered a 
lower-limit, flagged as such in the catalog, and set to 
$HR = (H - S_{\rm limit})/(H + S_{\rm limit})$, where 
$S_{\rm limit} = 2~\times$ soft-band background counts.

Column~8 contains an optical flag code essential to interpretation
of the optical data: 0 = no optical coverage, 1 = a solid optical ID,
2 = no optical counterpart (the magnitude listed is then a lower limit),
3 = saturated in the SEXSI optical image ($R$-band magnitude taken
from the Guide Star Catalog II~\citep{McLean:00} or a secondary
source in VIZIER database), 4 = a solid ID but
$R$-band magnitude affected by nearby bright source, 5 = more than one
optical source in X-ray error circle (the source with the smallest
positional offset is recorded in the table), and 6 = lower limit (no optical
counterpart) but area contaminated by nearby bright source so limiting
magnitude is unknown. Column~9 is the $R$-band magnitude of the optical
counterpart, with its error shown in column~10. Column~11 gives the
limiting magnitude for the image from which the optical counterpart,
or a lower limit thereto, was derived.  Column~12 lists $\Delta\alpha \equiv 
\alpha_{\rm x} - \alpha_{\rm o}$ in arcsec, where $\alpha_{\rm x}$ is the
astrometrically-corrected X-ray source RA (from column 2) and $\alpha_{\rm
o}$ is the optical counterpart RA; column 13 is the corresponding value
for declination.  Column 14, $\Delta r$ ($\equiv \sqrt{\Delta\alpha^{2}
+ \Delta\delta^{2}}$), is the X-ray to optical position difference, also
in arcsec.  Following \citet{Hornschemeier:01} and \citet{Stern:02b},
the logarithmic X-ray-to-optical flux ratio (column~15) is given by the
relation \begin{equation} \log{(f_{\rm x}/f_{\rm o})} = \log{f_{\rm x}}
+ (R/2.5) + 5.50, \end{equation} derived from the Kron-Cousins $R$-band
filter transmission function.  

Column 16 presents the SExtractor CLASS\_STAR parameter \citep{Bertin:96}
for basic star-galaxy separation, with values
ranging from 0.0 for significantly extended sources to 1.0 for 
sources consistent with perfectly stellar PSFs. 
The stellarity value of the 8 bright
sources with optflag = 3 (from the literature) is 
set to 1.00. 
We present the stellarity data 
with the caveat that the values should only be used for broad 
separation, for example, of sources near 0 verses 1, but not for 
detailed quantitative analysis.

Note that for all of the optical counterpart data columns, special
attention must be paid to the optical flag. For example, if the code is
a 2 (optical counterpart is not detected), column~9 describes a lower
limit, columns~10, 12 -- 14, and 16 have no data, and the X-ray-to-optical
flux ratio in column 15 is a lower limit.

\section{Discussion}
\label{sec:results}

We now discuss the results of the optical identifications of
the hard X-ray source  counterparts
and how the counterpart properties relate to the X-ray properties.
Throughout this section, only sources with a solid optical counterpart or
upper limit thereto are plotted and analyzed; we exclude sources with $R$-band
magnitudes or limits contaminated by a nearby bright source and sources
with more than one optical source within the identification radius.
This provides a sample of 947 hard X-ray sources with unambiguous counterpart
photometry, of which 603 are identifications and 344 have limits to the
counterpart optical magnitude. 

\begin{landscape}
\begin{figure}
\epsscale{0.90} 
\plotone{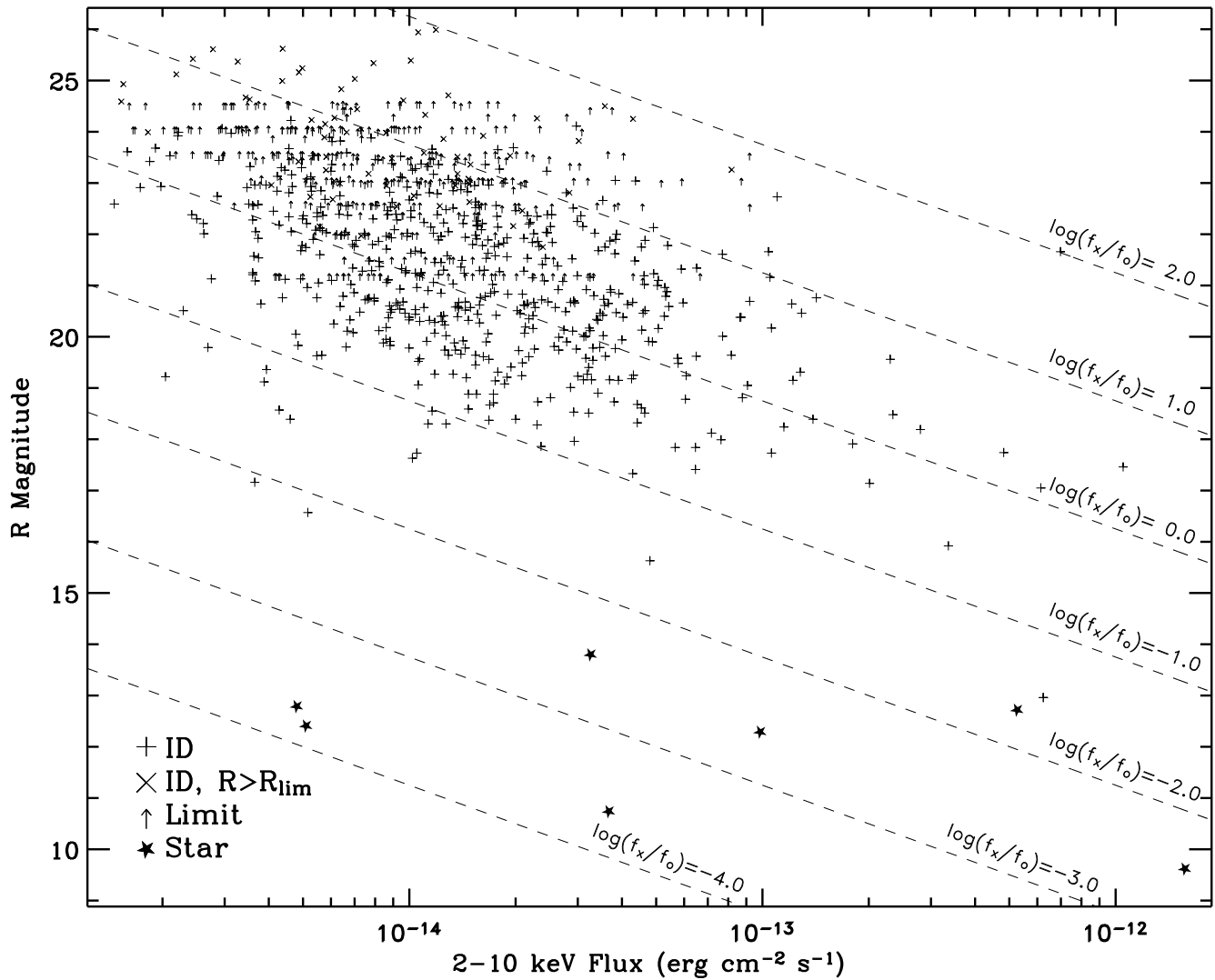} 
\caption{Optical magnitude of SEXSI sources, plotted as a function of
their hard-band (\hardrange) X-ray flux.  Dashed, angled lines show
constant $\log{(f_{\rm x}/f_{\rm o})}$.  Plus signs show sources with
optical counterparts, crosses are sources with a SExtractor ID but 
where $R > R_{\rm limit}$, while arrows denote sources lacking counterparts
to the image $R$-band limit.  Star symbols show SEXSI sources identified as
stars in the literature (only sources with $\log{(f_{\rm x}/f_{\rm o})} 
< -1$ were checked).}
\label{fig:r_hflux_scatter}
\end{figure}
\end{landscape}

Figure~\ref{fig:r_hflux_scatter} presents the $R$-band magnitudes,
or limits thereto, plotted as a function of \hardrange\ X-ray flux.  
Dashed lines show
constant X-ray-to-optical flux ratios.  The bulk of the SEXSI sources
have $-1 < \fxfo < 1$ and are fainter than $R = 20$.  Shallow, wide-area
X-ray surveys, such as the {\em ASCA} hard X-ray survey reported by
\citet{Akiyama:00}, find that the majority of extremely bright X-ray
sources ($S_{\rm 2-10~keV} \simgt 10^{-13} \fluxu$) are AGN and have $-1
< \fxfo < 1$.  Many of the SEXSI sources are likely fainter and/or more
distant analogs.

As has been found in other surveys \citep[e.g.,][]{Hornschemeier:01,
Stern:02b}, SEXSI detects a population of sources overluminous in the
X-ray for their optical magnitudes.  This population is found over the
entire X-ray flux range sampled.  Two explanations have been commonly
discussed to explain large values of $\fxfo$:  (i) extremely
high-redshift AGNs that might have bright X-ray fluxes but faint
$R$-band fluxes due to absorption from the Lyman transitions of
hydrogen along our line of sight, or (ii) heavy obscuration by material
in the host galaxy.  The latter interpretation for the majority of
sources is supported by near-IR studies, which find a large fraction to
be very red early-type galaxies at $z \sim 1$ \citep{Mignoli:04}.  In
addition, a number of {\em Type~II quasars}, which would have similar
properties, have been identified in the deepest {\em Chandra} surveys
\citep[e.g.,][]{Norman:02, Stern:02a, Dawson:03}.

By virtue of the large area surveyed, SEXSI contains numerous sources
with $\fxfo > 1.0$ at moderate X-ray fluxes, $S_{\rm 2-10~keV} \simgt
10^{-14} \fluxu$.  Many of these sources have $R \simlt 24$, well
within the reach of 8- to 10-meter class telescopes for spectroscopic
follow-up.  These are brighter examples of the new X-ray overluminous
population.  Deep Survey versions of these sources are typically
extremely faint --- as an example, \citet{Koekemoer:04} report on seven
``extreme X-ray / optical'' (EXO) sources in the 2~Msec {\em Chandra}
Deep Field-North (CDF-N) which, despite having extremely robust
detections in the {\em Chandra} data ($25 - 89$ counts), remain
undetected ($z_{850} > 27.9, 3\sigma$) in the Great Observatories
Origins Deep Survey imaging of the CDF-N \citep{Giavalisco:04}.  
\citet{Fiore:03}~report on the spectra of 13 sources with 
$\fxfo > 1$ from the HELLAS2XMM survey. 
They find 8 narrow-lined sources, with $L_{\rm x} > 10^{44}$ erg s$^{-1}$.
Our survey contains 109 such sources with $\fxfo > 1$ of 
which 64 have an optical limiting magnitude and $\simgt 20$ have spectral data.
These high $f_{\rm x}/f_{\rm o}$ sources provide a useful catalog for 
future infrared surveys. 

We also find a number of sources with very bright ($R < 14$) optical
counterparts.  The bulk of these sources 
(7 out of 8) are identified as stars in the literature. 

\begin{figure}
\epsscale{1} 
\plotone{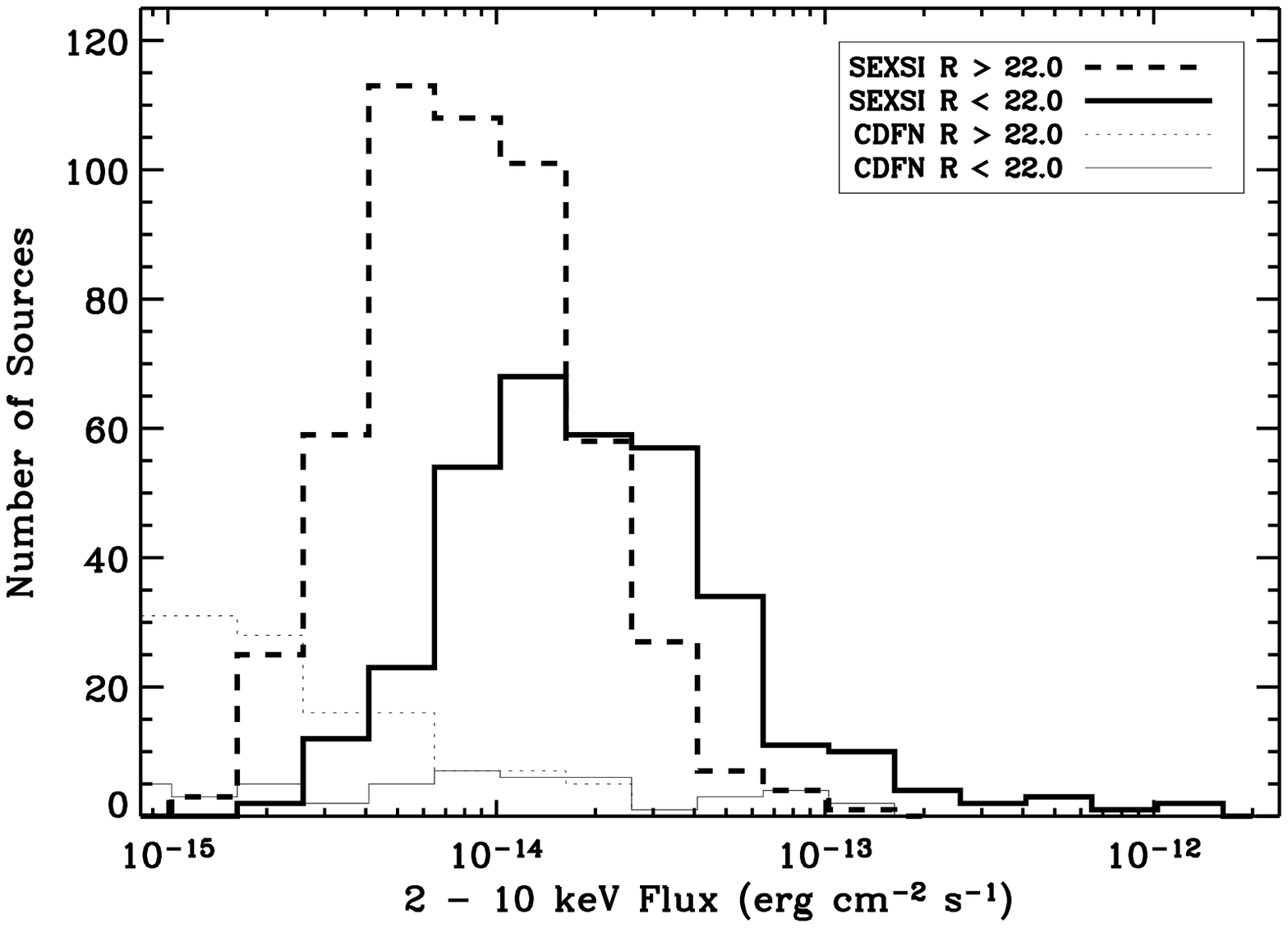} 
\caption{Histogram of \hardrange~fluxes, split by optical counterpart
$R$-band magnitude.  Heavy lines show SEXSI sources; lighter lines
show sources from the CDF-N \citep{Alexander:03, Barger:03}.  For both,
solid lines refer to optically-brighter sources ($R < 22.0$),
while dashed lines refer to optically-fainter sources ($R > 22.0$).
CDF-N X-ray fluxes have been converted from the published 2 -- 8 keV values to
the \hardrange~fluxes plotted here.  This involved converting X-ray photon
indices $\Gamma$
from the individually-derived indices used in the CDF-N X-ray catalog 
\citep{Alexander:03}~
to the average $\Gamma = 1.5$ adopted by the SEXSI project.
 In addition, the sources with $R_{\rm limit} < 22.0$ were excluded from the plot.}
\label{fig:hfluxhist} 
\end{figure}

\begin{figure}
\epsscale{1} 
\plotone{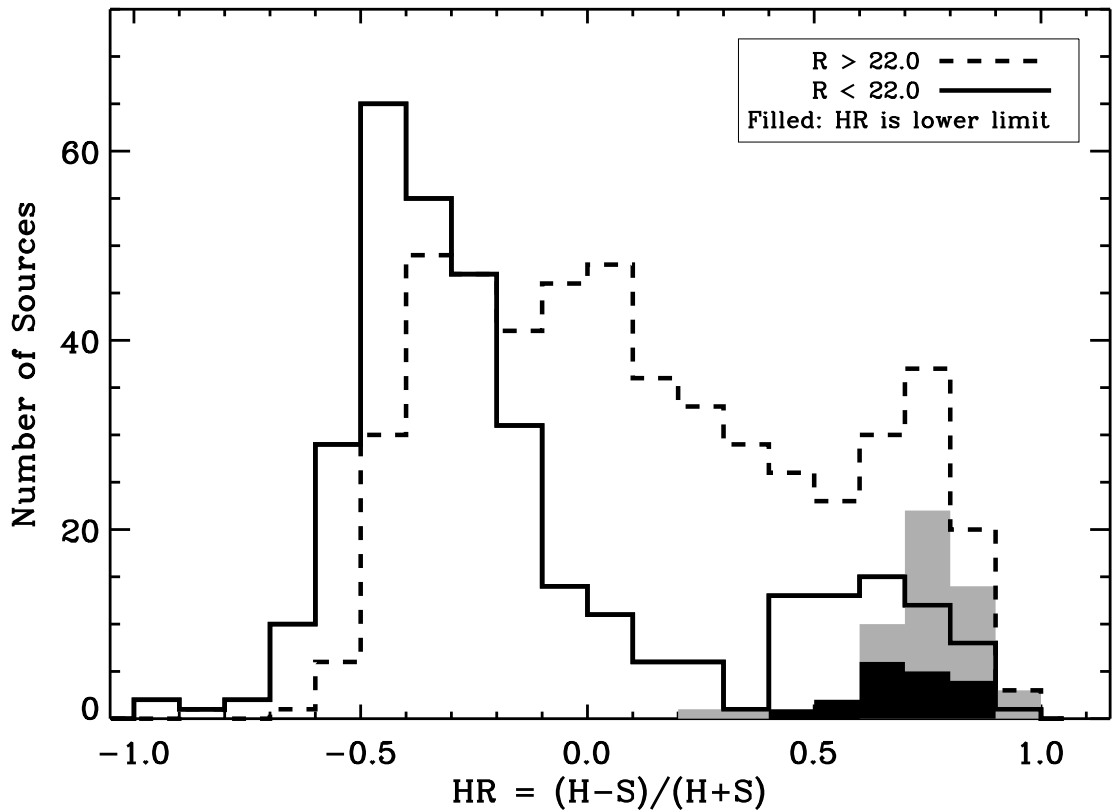} 
\caption{Histogram of X-ray hardness ratios ($HR$), split by optical
counterpart $R$-band magnitude.  The solid line shows optically-brighter
($R < 22.0$) sources, while the dashed line shows optically-fainter ($R >
22.0$) sources; objects with limiting $HR$s are included in these histograms,
plotted at the limiting $HR$. The filled histograms present only the 
sources with limiting $HR$ for $R >22$ (gray) and $R < 22$ (black).}

\label{fig:hrhist}
\end{figure}

\begin{figure}
\epsscale{1} 
\plotone{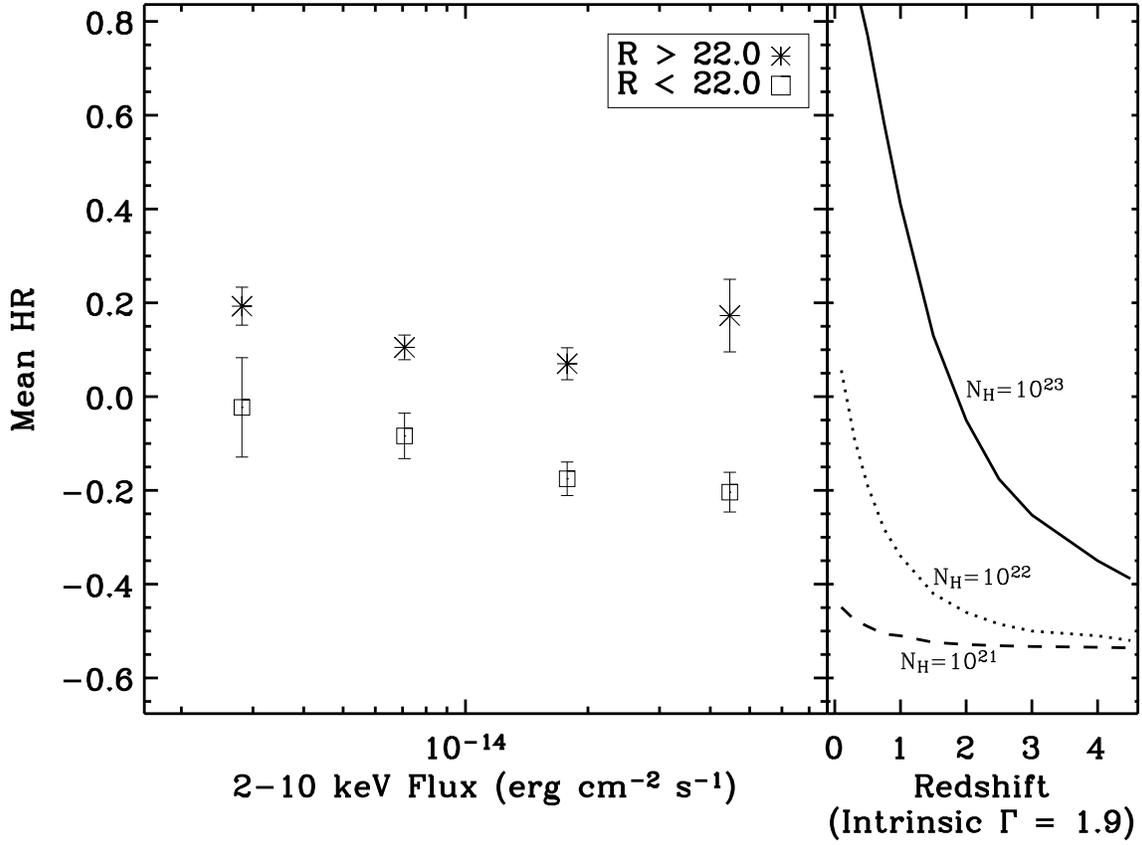} 
\caption{Mean hardness ratio ($HR$) of sources split by optical
counterpart $R$-band magnitude, as a function of \hardrange~flux.
Sources were split at $R = 22.0$ and then binned into four equally spaced
logarithmic X-ray flux bins.  Note that at each X-ray flux, the optically
faint sources (asterisks) are significantly harder than the brighter
optical sources (squares). The right panel shows $HR$ as a function of 
redshift for an intrinsic $\Gamma=1.9$ spectrum with several 
intrinsic obscuring column densities, for reference.}

\label{fig:meanhr} 
\end{figure}

\begin{landscape}
\begin{figure}
\epsscale{0.9} 
\plotone{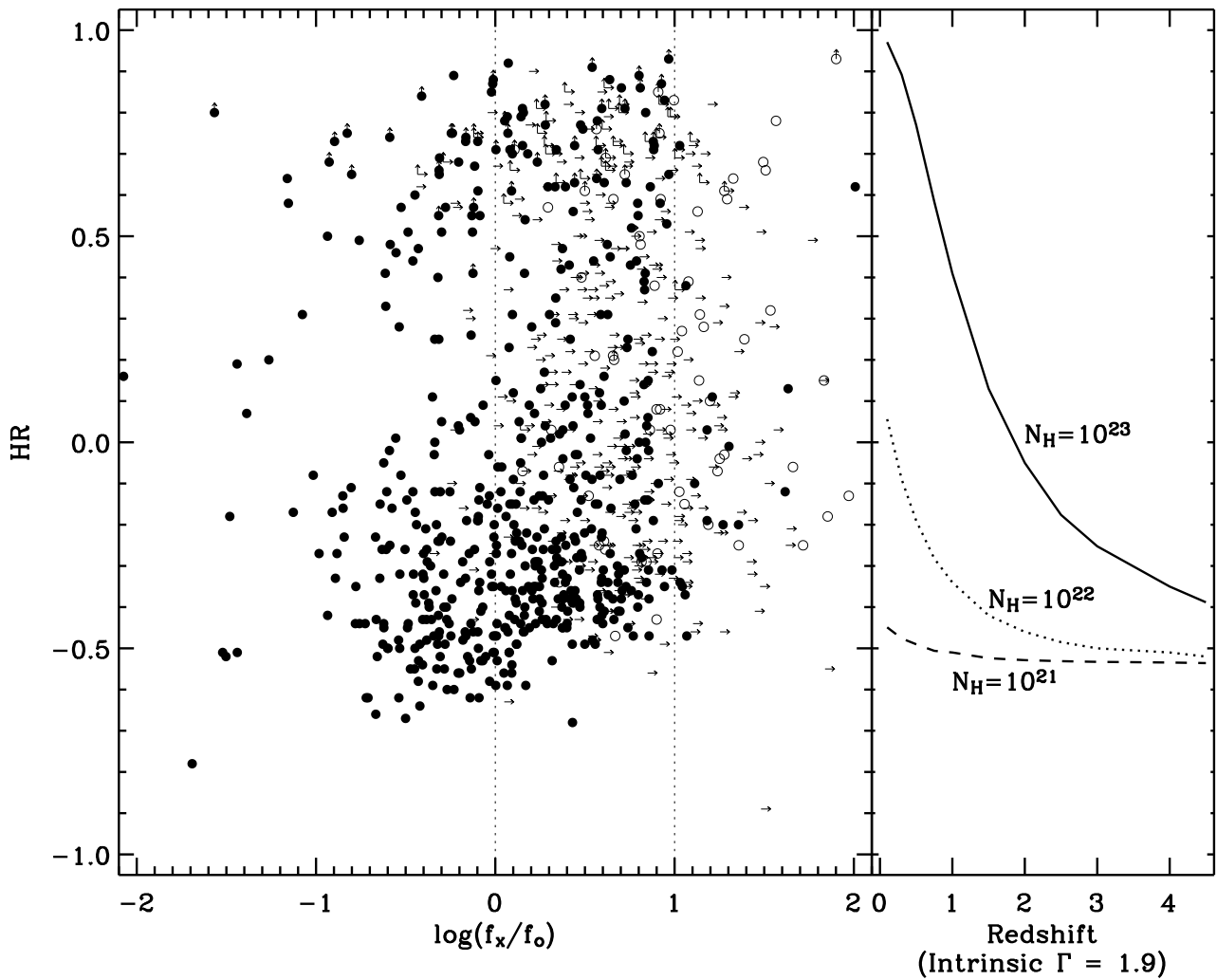} 
\caption{Hardness ratio as a function of X-ray-to-optical flux ratio. 
Sources with identified optical counterparts are shown with filled
($R < R_{\rm limit}$) and open ($R > R_{\rm limit}$) circles;
sources with an optical limit are shown as right-pointing arrows 
-- the $f_{\rm x}/f_{\rm o}$ 
value is a lower limit. Vertical arrows indicate a lower limit to the 
$HR$.  The right panel shows $HR$ as a function of 
redshift for an intrinsic $\Gamma=1.9$ spectrum with several 
intrinsic obscuring column densities, for reference.}
\label{fig:fxfohr} 
\end{figure}
\end{landscape}

Figure~\ref{fig:hfluxhist} presents a histogram of the number 
of sources in SEXSI and in the
CDF-N~\citep{Alexander:03,Barger:03} as a function of hard
X-ray flux and split by $R$-band magnitude at $R=22$. This 
magnitude was chosen
to separate our sources approximately in half.
The small number of sources from images with $R_{\rm limit} <22$ 
were not included in this plot or in Figures~\ref{fig:hrhist},
~\ref{fig:meanhr}, and~\ref{fig:fxfohr}.
The majority of the SEXSI
sources have $S_{\rm 2-10~keV} \approx 10^{-14} \fluxu$, a flux level
that lies between the {\em ASCA} and {\em BeppoSAX} sensitivity limits
and the {\em Chandra} Deep Survey capability.  This important flux
range corresponds to the regime in which the \lognlogs~relation changes slope
and from which the bulk of the \hardrange~X-ray background arises
\citep{Cowie:02, Harrison:03}.  Compared to the Deep
Surveys, SEXSI has nearly an order of magnitude more sources at this
X-ray depth, approximately half of which have $R < 22$, making them easy 
spectroscopic targets for 8- to 10-meter class telescopes.

Figure~\ref{fig:hrhist} shows the hardness ratio histogram of the SEXSI
sources, again split at $R=22$. The optically-brighter sources are
peaked at a low hardness ratio, while the fainter sources have a much
harder, and broader, distribution, showing that the optically-fainter
portion of the sample has a higher fraction of sources with a flat
X-ray spectral slope. The large peak of the $R<22$ sources is near $HR \sim
-0.5$, which corresponds to a power-law photon index ($\Gamma$) between
1.8 and 1.9, a typical value for unobscured broad-line AGN.  This
suggests that the majority of the optically-brighter sources are
broad-lined AGN.  Spectroscopic followup (Paper III) will be able to
specifically address this hypothesis. The harder and broader
distribution of the $R>22$ sources indicates that obscuration at the
source is likely involved.  \citet{Alexander:01}~shows a similar
trend in the 1 Ms CDF-N data.

Figure~\ref{fig:meanhr}, which displays the mean hardness ratio as a
function of hard-band flux separately for bright and faint counterpart
magnitudes, again emphasizes that distinct populations are contributing
to the overall source counts. 
The mean values for this figure are calculated using source $HR$s
regardless of if they are flagged as lower limits. Setting all such 
flagged sources to the upper extreme, $HR=1$, shifts the points only
slightly, if at all.
For reference, the $HR$ of a $\Gamma=1.9$ spectrum 
with varying intrinsic absorbing column density and redshift is
presented in the adjacent panel. This shows that modest-redshift,
heavily-obscured sources produce the highest $HR$s.
For sources fainter than $R=22$, the mean
hardness ratio is essentially constant at $HR=+0.15$ over two orders of
magnitude in X-ray flux. For those sources with brighter magnitudes,
however, there is a significant decline in mean hardness ratio, from
$HR \sim -0.25$ at $\hardfluxrange \sim 10^{-13} \fluxu$ to $HR \sim
-0.1$ at $\hardfluxrange \sim 10^{-15} \fluxu$. In a simple-minded
two-component model with median hardness ratios of $-0.35$ and $+0.2$
for the two source classes, the faint sources are dominated by the hard
population with perhaps an admixture of 10\% soft sources at all flux
levels, while for the brighter sources, the ratio of hard to soft
sources changes from 20:80 at $\hardfluxrange \sim 10^{-13} \fluxu$ to
80:20 at $\hardfluxrange \sim 10^{-15} \fluxu$.  This is qualitatively
consistent with the \lognlogs\ relations derived in Paper I, where we
showed that hard sources follow a constant power-law over three orders
of magnitude in hard-band flux, while softer sources show a distinct
break at $\hardfluxrange \sim 10^{-14} \fluxu$. In this picture, the
soft sources are the standard, largely unobscured AGN at all redshifts,
while the harder component represents the lower mean-redshift,
lower-luminosity populations revealed in the Deep Surveys
\citep[e.g.,][]{Tozzi:01,Barger:02}.

Figure~\ref{fig:fxfohr} shows the hardness ratio as a function of X-ray to optical 
flux ratio for the SEXSI sample, again with the $HR$ of a $\Gamma=1.9$ spectrum 
with varying intrinsic absorbing column density as a function of redshift presented
 in the adjacent panel. The main plot shows a trend toward
larger hardness ratios as $f_{\rm x}/f_{\rm o}$ increases.
This is in general consistent with the notion that 
the increasing $f_{\rm x}/f_{\rm o}$ 
results from larger absorbing columns which attenuate the 
$R$-band flux much more severely than the hard X-ray flux 
\citep[e.g.,][]{Mainieri:02}. 
It is interesting,
however, that the most extreme sources with $\fxfo > 1$ are not all 
hard, but exhibit a wide range of $HR$s, implying that obscuration may 
not be the sole explanation for the dim optical counterparts.
\citet{Fiore:03}~find that for non-broad-lined sources (e.g., 
sources with optical spectra showing narrow AGN 
emission lines or early-type galaxy absorption lines) 
there is a linear correlation between $\fxfo$
and $\log(L_{\rm 2 - 10~keV})$.
 This relationship implies
that some of the high $f_{\rm x}/f_{\rm o}$ sources with low $HR$s are
 high luminosity, high redshift, narrow-lined AGN.

We also explored the optical source densities in the vicinity of the X-ray
sources. For each optical counterpart, we counted the number of
sources in the surrounding region and compared the source density to the 
overall field density to calculate overdensities. 
We employed 20\arcsec, 30\arcsec, 40\arcsec, and 60\arcsec\ radius circles
around each optical counterpart and counted sources with
$R$-magnitude of $|R - R_{\rm counterpart}| < 1$~and 
$|R - R_{\rm counterpart}| < 2$~for each of the sources. 
We found no significant overdensities around the 
X-ray sources.

\section{Summary}
\label{sec:summary}

We present $R$-band imaging of~$\sim 95$\% of the \hardrange\ X-ray
sources in the SEXSI survey. We describe our optical data reduction and
the X-ray to optical source matching algorithms employed, and present a
catalog of $R$-band counterpart photometry and astrometry.  While the
power of the SEXSI sample is fully realized only with the addition of
our collection of optical counterpart spectra, the photometric
identification of the sources is an important step that provides clues
to the composition of the source population. We find that by splitting
the sources at $R=22$ and analyzing each group's X-ray properties, we
begin to see the emergence of what appear to be physically distinct
populations. For example we find that the hardness ratio distribution of
the optically brighter sources is sharply peaked near $HR \sim -0.5$,
typical of unobscured AGN, while the fainter, $R>22$ sources have a
much broader and harder spectral distribution.  The findings of this
paper suggest that the analysis of optical spectral data from
counterparts fainter than $R=22$ (attainable with 10-meter-class
telescopes), will be an essential part of exploring the X-ray
background composition.

\acknowledgements

This research has made use of the NASA/IPAC Extragalactic Database
(NED) which is operated by the Jet Propulsion Laboratory, California
Institute of Technology, under contract with the National Aeronautics
and Space Administration (NASA).  Some of the data presented herein
were obtained at the W.M. Keck Observatory, which is operated as a
scientific partnership among the California Institute of Technology,
the University of California and NASA. The Observatory was made possible
by the generous financial support of the W.M. Keck Foundation. The
authors wish to recognize and acknowledge the very significant cultural
role and reverence that the summit of Mauna Kea has always had within
the indigenous Hawaiian community.  We are most fortunate to have the
opportunity to conduct observations from this mountain.  This work has
made use of the USNOFS Image and Catalog Archive operated by the United
States Naval Observatory, Flagstaff Station.  
The Guide Star Catalog was produced at the Space Telescope Science 
Institute under U.S. Government grant with data based on photographic 
data obtained using the Oschin Schmidt Telescope on Palomar Mountain and 
the UK Schmidt Telescope. This research has made use of NASA's 
Astrophysics Data System.
This work has been supported
by NASA NAG5-6035 (DJH), as well as by a small {\em Chandra} archival
grant.  The work of DS was carried out at the Jet Propulsion Laboratory,
California Institute of Technology, under a contract with NASA.

\clearpage

\begin{deluxetable}{llccl}
\tablecaption{Optical imaging instruments employed in the SEXSI survey.}
\tablecolumns{4}
\tablewidth{0pc}
\tablehead{& & \colhead{Plate Scale} & & \\
\colhead{Camera} & \colhead{Telescope} & [\arcsec /pix] & FOV & \colhead{Reference}}
\startdata
8K	      & MDM 2.4m      & 0.18	& $24.6\arcmin \times 24.6\arcmin$ & \citet{Crotts:01} \\
CCD13	      & Palomar 60''  & 0.37	& $12.6\arcmin \times 12.6\arcmin$ & \\
Cosmic	      & Palomar 200'' & 0.29	& $ 9.9\arcmin \times  9.9\arcmin$ & \citet{Kells:98} \\
Echelle       & MDM 1.3m      & 0.50	& $17.1\arcmin \times 17.1\arcmin$ & \\
Echelle	      & MDM 2.4m      & 0.28	& $ 9.6\arcmin \times  9.6\arcmin$ & \\
LFC	      & Palomar 200'' & 0.18	& $\sim \pi (12.3\arcmin)^2$       & \citet{Simcoe:00} \\
LRIS	      & Keck I	      & 0.22	& $ 7.5\arcmin \times  7.5\arcmin$ & \citet{Oke:95} \\
Templeton     & MDM 1.3m      & 0.50	& $ 8.5\arcmin \times  8.5\arcmin$ & \\
\enddata
\label{tbl:cameras}
\end{deluxetable}

\begin{deluxetable}{lcccclllcc}
\tabletypesize{\footnotesize}
\tablecaption{SEXSI optical imaging.}
\tablecolumns{10}
\tablewidth{0pc}
\tablehead{& \colhead{X-ray} & & & & \multicolumn{5}{c}{Optical Images} \\
\cline{6-10}
 & \colhead{Exp.} & \colhead{$\langle \Delta\alpha_{\rm xo} \rangle$} & \colhead{$\langle \Delta\delta_{\rm xo} \rangle$} & \colhead{extinction\tablenotemark{a}} &\colhead{RA} &\colhead{DEC} & Camera & \colhead{$R$ limit\tablenotemark{b}} & \colhead{Seeing} \\ 
\colhead{Target} & \colhead{[ks]} & \colhead{[\arcsec]} &  \colhead{[\arcsec]} & \colhead{[$R$ Mag]} & \colhead{(J2000)} & \colhead{(J2000)} &  & \colhead{[Mag]} &\colhead{[\arcsec]}}
\startdata
NGC 891      &  51 & -0.9 $\pm$ 0.3  & ~0.6 $\pm$ 0.2 & 0.17                  & 02 22 40 & +42 26 16 & 8K & 24.3 & 1.6 \\
AWM 7        &  48 & ~0.5 $\pm$ 0.3  & ~0.2 $\pm$ 0.1  & 0.30                  & 02 54 45 & +41 40 10 & 8K & 24.2\tablenotemark{c} & 1.7\\
XRF 011130   &  30 & ~0.4 $\pm$ 0.2  & -0.5 $\pm$ 0.2 & 0.26                  & 03 05 28 & +03 48 52 & LFC & 23.2\tablenotemark{d} & 1.2 \\
NGC 1569     &  97 & -0.3 $\pm$ 0.2  & ~0.6 $\pm$ 0.3 & 2.00\tablenotemark{e} & 04 29 08 & +64 45 42 & Echelle (1.3) & 21.5 & 1.7 \\
 &  &  &  &  & 04 29 42 & +64 42 35 & LRIS  & 21.5 & 0.9 \\
 &  &  &  &  & 04 29 53 & +64 39 47 & CCD13 & 21.0 & 1.4 \\
 &  &  &  &  & 04 30 36 & +64 51 06 & LRIS  & 23.0 & 1.2 \\
 &  &  &  &  & 04 31 15 & +64 51 06 & LRIS  & 22.5 & 0.9 \\
3C 123       &  47 & -0.1 $\pm$ 0.1 & ~0.0 $\pm$ 0.0 & 2.61\tablenotemark{e} & 04 36 52 & +29 40 08 & 8K & 21.9 & 1.4 \\
CL 0442+0202\tablenotemark{f} &  44 &  -0.2 $\pm$ 0.2  &  -1.2 $\pm$ 0.3 & 0.41                  & 04 42 17 & +02 03 35 & LFC & 21.1\tablenotemark{d} & 1.2 \\
CL 0848+4454 & 186 & -0.3 $\pm$ 0.1 & -1.2 $\pm$ 0.1 & 0.08                  & 04 48 49 & +44 54 09 & LFC & 24.4\tablenotemark{d} & 1.1 \\
RX J0910     & 171 & -0.3 $\pm$ 0.1 & -2.3 $\pm$ 0.2 & 0.05                  & 09 10 33 & +54 19 38 & 8K & 24.0 & 1.6 \\
1156+295     &  49 & -0.5 $\pm$ 0.2 & -1.1 $\pm$ 0.3 & 0.05                  & 11 59 32 & +29 16 19 & 8K & 24.5 & 1.5 \\
NGC 4244     &  49 & -0.2 $\pm$ 0.2 & ~0.0 $\pm$ 0.1 & 0.06                  & 12 16 46 & +37 51 46 & CCD13 & 22.9\tablenotemark{d} & 1.8 \\
 &  &  &  &  & 12 17 34 & +37 46 54 & Echelle (1.3)&  23.4\tablenotemark{d}  & 1.8 \\
 &  &  &  &  & 12 17 54 & +37 55 15 & CCD13 & 23.4\tablenotemark{d} & 1.5 \\
NGC 4631     &  59 & ~0.4 $\pm$ 0.3 & -0.3 $\pm$ 0.2 & 0.05                  & 12 41 29 & +32 36 46 & CCD13 & 23.0\tablenotemark{d} & 1.6 \\
 &  &  &  &  & 12 41 39 & +32 30 44 & CCD13 & 23.0\tablenotemark{d} & 2.5 \\
 &  &  &  &  & 12 42 03 & +32 40 53 & Echelle (2.4)& 23.0\tablenotemark{d} & 1.2  \\
 &  &  &  &  & 12 42 21 & +32 37 09 & CCD13 & 23.0\tablenotemark{d} & 1.7 \\
 &  &  &  &  & 12 42 59 & +32 32 49 & Templeton  & 23.0\tablenotemark{d} & 1.4 \\
HCG 62       &  49 & ~0.1 $\pm$ 0.1 & -0.1 $\pm$ 0.1 & 0.13                  & 12 53 04 & $-$09 14 20 & Echelle (2.4) & 22.4\tablenotemark{d} & 1.2 \\
 &  &  &  &  & 12 53 06  & $-$09 05 48 & Echelle (2.4)& 23.4\tablenotemark{d} & 1.0  \\
 &  &  &  &  & 12 53 04 & $-$09 14 20 & CCD13 & 22.4\tablenotemark{d} & 1.6 \\
RX J1317     & 111 & -1.6 $\pm$ 0.2 & -1.6 $\pm$ 0.2 & 0.03                  & 13 17 13 & +29 11 32 & LFC & 23.5\tablenotemark{d} & 1.5 \\
BD 1338      &  38 & -1.3 $\pm$ 0.2 & -1.1 $\pm$ 0.1 & 0.04                  & 13 37 44 & +29 25 24 & CCD13 & 22.5 & 1.3 \\
 &  &  &  &  & 13 37 53 & +29 29 44 & CCD13 & 22.5 & 1.7 \\
 &  &  &  &  & 13 38 15 & +29 36 51 & CCD13 & 22.5 & 2.0 \\
 &  &  &  &  & 13 38 26 & +29 24 53 & CCD13 & 22.5 & 1.5 \\
RX J1350     &  58 & ~0.7 $\pm$ 0.3 & -2.5 $\pm$ 0.5 & 0.03                  & 13 50 39 & +60 04 11 & LFC & 22.5\tablenotemark{g} & 0.9 \\
3C 295       &  23 & ~0.2 $\pm$ 0.2 & -0.3 $\pm$ 0.3 & 0.05                  & 14 11 01 & +52 21 18 & CCD13 & 22.5\tablenotemark{d} & 2.2 \\
 &  &  &  &  & 14 11 24 & +52 13 36 & CCD13 & 21.5\tablenotemark{d} & 2.8 \\
 &  &  &  &  & 14 11 46 & +52 05 42 & CCD13 & 22.0\tablenotemark{d} & 2.3 \\
GRB 010222   &  18 & -0.6 $\pm$ 0.2 & ~0.1 $\pm$ 0.1 & 0.06                  & 14 51 59 & +43 08 54 & CCD13 & 21.9 & 2.1 \\
 &  &  &  &  & 14 52 13 & +43 01 06 & CCD13 & 22.9 & 1.4 \\
 &  &  &  &  & 14 52 43 & +42 55 12 & CCD13 & 22.4 & 1.6 \\
 &  &  &  &  & 14 52 59 & +43 06 07 & CCD13 & 22.9 & 1.3 \\
QSO 1508     &  89 & -0.6 $\pm$ 0.2 & -0.2 $\pm$ 0.1 & 0.03                  & 15 09 50 & +57 04 16 & LFC & 24.0\tablenotemark{g} & 1.2 \\
MKW 3S       &  57 & -0.8 $\pm$ 0.1 & ~0.1 $\pm$ 0.1 & 0.09  & 15 21 30 & +07 47 43 & CCD13 & 23.4 & 1.2 \\
 &  &  &  &  & 15 21 37 & +07 39 08 & CCD13 & 22.9 & 1.1 \\
 &  &  &  &  & 15 22 12 & +07 48 36 & CCD13 & 22.9 & 1.3 \\
 &  &  &  &  & 15 22 13 & +07 47 21 & CCD13 & 22.9 & 1.6  \\
MS 1621      &  30 & -1.2 $\pm$ 0.1 & ~0.3 $\pm$ 0.2 & 0.09                  & 16 23 06 & +26 36 33 & CCD13 & 22.4 & 2.3 \\
 &  &  &  &  & 16 23 13 & +26 30 22 & Cosmic & 24.4 & 1.4  \\
 &  &  &  &  & 16 23 26 & +26 38 28 & Cosmic & 23.9 & 1.6 \\
 &  &  &  &  & 16 23 42 & +26 43 49 & CCD13 & 22.9 & 2.3 \\ 
 &  &  &  &  & 16 24 03 & +26 35 32 & Cosmic & 23.9 & 1.2 \\
GRB 000926   &  32 & -0.7 $\pm$ 0.2 & -0.6 $\pm$ 0.3 & 0.08                  & 17 03 19 & +51 47 58 & 8K & 22.9 & 1.2 \\
RX J1716     &  52 & -0.5 $\pm$ 0.1 & ~1.2 $\pm$ 0.1 & 0.09                  & 17 15 36 & +67 19 45 & CCD13 & 22.4 & 1.7 \\
 &  &  &  &  & 17 15 55 & +67 05 59 & Cosmic & 24.4 & 1.3 \\
 &  &  &  &  & 17 16 39 & +67 13 27 & Cosmic & 23.9 & 1.4 \\
 &  &  &  &  & 17 17 11 & +67 01 40 & Cosmic & 24.4 & 1.2 \\
 &  &  &  &  & 17 17 42 & +67 12 36 & CCD13 & 22.9 & 1.3 \\
 &  &  &  &  & 17 17 56 & +67 09 08 & Cosmic & 23.9 & 1.6 \\
 &  &  &  &  & 17 18 04 & +67 12 01 & CCD13 & 22.4 & 2.2 \\
NGC 6543  &  46 & -0.5 $\pm$ 0.2 & -0.2 $\pm$ 0.1 & 0.12                  & 17 57 44 & +66 44 40 & CCD13 & 22.9 & 1.3 \\
 &  &  &  &  & 17 58 31 & +66 34 02 & CCD13 & 22.9 & 1.4 \\
 &  &  &  &  & 18 00 25 & +66 30 52 & CCD13 & 22.9 & 1.2 \\
XRF 011030   &  47 & -0.2 $\pm$ 0.2 & -0.4 $\pm$ 0.2 & 1.09\tablenotemark{e} & 20 44 02 & +77 21 06 & LFC & 21.4\tablenotemark{d} & 2.2 \\
MS 2053      &  44 & -1.1 $\pm$ 0.1 & ~0.4 $\pm$ 0.1 & 0.22                  & 20 55 48 & $-$04 31 10 & CCD13 & 22.8 & 1.7 \\
 &  &  &  &  & 20 55 59 & $-$04 35 47 & Cosmic & 23.8 & 1.4 \\
 &  &  &  &  & 20 56 09 & $-$04 44 01 & Cosmic & 23.8 & 1.4 \\
 &  &  &  &  & 20 56 30 & $-$04 38 17 & LRIS & 23.8 & 1.0 \\
 &  &  &  &  & 20 56 33 & $-$04 30 53 & CCD13 & 22.3 & 2.1 \\
 &  &  &  &  & 20 56 42 & $-$04 41 28 & Cosmic & 23.8 & 1.2 \\
RX J2247     &  49 & -0.8 $\pm$ 0.1 & ~1.0 $\pm$ 0.1 & 0.15                  & 22 47 21 & +03 39 25& 8K & 23.4 & 1.7 \\
Q2345        &  74 & ~0.0 $\pm$ 0.1 & -0.3 $\pm$ 0.2 & 0.07                  & 23 48 25 & +01 01 12& 8K & 23.4\tablenotemark{d} & 1.5 \\
\enddata
\tablenotetext{a} {Error on Galactic $R$-band extinction is $\sim 0.02$, with exceptions noted.
This error is generally an order of magnitude smaller than the error on the zeropoint magnitude.}
\tablenotetext{b} {Image $R$-band limits are Galactic extinction-subtracted and correct to within 0.5 mag.}
\tablenotetext{c} {Limiting magnitude estimated from visual inspection;
the nearby cluster sources skew the $R$-band number counts distribution
and gave an unreasonable limiting magnitude from our automated algorithm.}
\tablenotetext{d} {Zeropoint derived using USNO-B1.0 catalog. No SDSS coverage available.}
\tablenotetext{e} {Error on Galactic $R$-band extinction is $\sim 0.2$.}
\tablenotetext{f} {In Paper I, Table 2, CL 0442+0202 was erroneously referred to as ``CL 0442+2200''.}
\tablenotetext{g} {Zeropoint derived using SDSS 1st data release.}
\label{tbl:images}
\end{deluxetable}

\begin{deluxetable}{lccrrrrclccrrcrrc}
\tablecaption{SEXSI optical counterpart catalog.}
\tabletypesize{\footnotesize}
\rotate
\tablecolumns{17}
\tablewidth{0pc}
\tablehead{\colhead{ }  & \multicolumn{6}{c}{\underline{~~~~~~~~~~~~~~~~~~~~~~~~~~~~~~X-ray Data~~~~~~~~~~~~~~~~~~~~~~~~~~~~~~}}&\multicolumn{9}{c}{\underline{~~~~~~~~~~~~~~~~~~~~~~~~Optical Counterpart Data~~~~~~~~~~~~~~~~~~~~~~~~}} \\ 
\colhead{CXOSEXSI\_} & \colhead{ } & \colhead{ } & \colhead{ } & \multicolumn{2}{c}{\underline{~~2 -- 10 keV~~}}& \colhead{} & \multicolumn{8}{c}{ } \\ 
\colhead{ } & \colhead{$\alpha_{\rm x}$ (J2000)}& \colhead{$\delta_{\rm x}$ (J2000)} & \colhead{OAA} & \colhead{Flux}& \colhead{SNR} & \colhead{$HR$} & \colhead{Flag} & \colhead{$R$}& \colhead{$\sigma_{\rm R}$} & \colhead{$R_{\rm limit}$}& 
\colhead{$\Delta \alpha$} & \colhead{$\Delta \delta$} & \colhead{$\Delta$r}& $\log{\frac{f_{\rm x}}{f_{\rm o}}}$ & Stel. \\
\colhead{(1)} & \colhead{(2)\tablenotemark{a}} & \colhead{(3)\tablenotemark{a}} & \colhead{(4)} & \colhead{(5)} & \colhead{(6)} & \colhead{(7)\tablenotemark{b}} & \colhead{(8)\tablenotemark{c}} & \colhead{(9)} & \colhead{(10)} & \colhead{(11)} & \colhead{(12)} & \colhead{(13)} & \colhead{(14)} & (15) & \colhead{(16)}}
\startdata
J022142.6+422654 &  02 21 42.75 & +42 26 53.5 &   9.49 &  23.10 &   2.83 &  -0.33  &  1  &  18.73  &  0.16   & 24.3 &  1.0 &  1.8 &  2.1 &             -0.64  &  0.66\\
J022143.6+421631 &  02 21 43.72 & +42 16 31.0 &   8.33 &  74.80 &   8.81 &  -0.28  &  6  &  ~\ldots &  \ldots & \ldots & \ldots &\ldots & \ldots& \ldots  &      	\ldots						\\
J022151.6+422319 &  02 21 51.76 & +42 23 18.7 &   6.17 &   6.55 &   2.06 &  -0.16  &  2  &  24.3  &  \ldots & 24.3 & \ldots &	\ldots & \ldots	&       1.04  &  \ldots	\\
J022204.9+422338 &  02 22 05.08 & +42 23 37.7 &   4.24 &   8.73 &   2.74 &   0.84\tablenotemark{d}  &  1  &  20.37  &  0.16   & 24.3 & -0.2 &  0.5 &  0.5 & 	      -0.41   &  0.02\\
J022205.1+422213 &  02 22 05.21 & +42 22 12.7 &   3.45 &   6.82 &   2.38 &   0.07  &  2  &  24.3  &  \ldots & 24.3 & \ldots &	\ldots & \ldots	&       1.07  &  \ldots	\\
J022207.1+422918 &  02 22 07.19 & +42 29 18.2 &   8.93 &  18.90 &   3.84 &  -0.33  &  1  &  21.56  &  0.16   & 24.3 &  2.3 & -0.8 &  2.4 &	       0.40   &  0.96\\
J022210.0+422956 &  02 22 10.08 & +42 29 55.7 &   9.38 &  17.30 &   3.52 &  -0.62  &  1  &  18.88  &  0.16   & 24.3 &  0.2 & -0.6 &  0.6 &	      -0.71   &  0.03\\
J022210.8+422016 &  02 22 10.93 & +42 20 16.1 &   2.17 &   3.63 &   1.53 &  -0.46  &  1  &  21.57  &  0.17   & 24.3 &  0.6 & -0.4 &  0.8 &	      -0.31   &  0.95\\
J022211.7+421910 &  02 22 11.79 & +42 19 10.1 &   2.55 &   6.87 &   2.36 &   0.14  &  6  &  ~\ldots &  \ldots & \ldots & \ldots &\ldots & \ldots& \ldots  &      	\ldots						\\
J022215.0+422341 &  02 22 15.12 & +42 23 41.0 &   3.15 &  15.40 &   3.88 &  -0.41  &  1  &  22.52  &  0.18   & 24.3 &  0.1 &  0.0 &  0.1 &	       0.70   &  0.64\\
J022215.1+422045 &  02 22 15.19 & +42 20 44.5 &   1.32 &  42.90 &   7.49 &  -0.42  &  1  &  17.33  &  0.16   & 24.3 & -0.3 &  0.1 &  0.3 &	      -0.94   &  0.03\\
J022215.5+421842 &  02 22 15.63 & +42 18 41.7 &   2.46 &   6.79 &   2.34 &   0.28  &  1  &  20.33  &  0.16   & 24.3 & -0.4 &  0.1 &  0.4 &	      -0.54   &  0.03\\
J022219.3+422052 &  02 22 19.40 & +42 20 51.6 &   0.54 &   6.38 &   2.31 &   0.37  &  1  &  23.82  &  0.22   & 24.3 & -0.9 &  0.6 &  1.1 &	       0.83   &  0.38\\
J022224.3+422138 &  02 22 24.45 & +42 21 38.4 &   0.91 &  64.60 &   8.92 &  -0.44  &  1  &  17.41  &  0.16   & 24.3 &  0.0 &  0.0 &  0.0 &	      -0.73   &  0.98\\
J022225.2+422451 &  02 22 25.33 & +42 24 50.9 &   4.07 & 128.00 &  12.99 &  -0.34  &  1  &  19.31  &  0.16   & 24.3 &  0.1 &  0.2 &  0.2 &	       0.33   &  0.98\\
J022226.5+422154 &  02 22 26.63 & +42 21 54.4 &   1.35 &   6.48 &   2.32 &  -0.35  &  2  &  24.3  &  \ldots & 24.3 & \ldots & \ldots & \ldots &        1.04  &   \ldots	\\
J022232.5+423015 &  02 22 32.61 & +42 30 14.6 &   9.61 &  53.60 &   6.50 &   0.12  &  1  &  20.88  &  0.16   & 24.3 &  0.4 &  0.4 &  0.5 &  	       0.58   &  0.77\\
J022236.3+421730 &  02 22 36.45 & +42 17 30.2 &   4.23 &  12.00 &   3.30 &   0.01  &  1  &  22.38  &  0.18   & 24.3 &  0.3 & -0.9 &  1.0 &  	       0.53   &  0.42\\
J022236.7+422858 &  02 22 36.88 & +42 28 57.7 &   8.57 &  12.90 &   3.00 &  -0.68  &  1  &  22.05  &  0.17   & 24.3 & -1.5 &  2.1 &  2.6 &  	       0.43   &  0.81\\
J022259.1+422434 &  02 22 59.18 & +42 24 33.6 &   7.77 &  32.50 &   5.43 &   0.49  &  2  &  24.3  &  \ldots & 24.3 & \ldots &\ldots & \ldots &         1.74  &   \ldots \\
J022334.0+422212 &  02 23 34.13 & +42 22 11.6 &  13.34 &  37.50 &   4.47 &   0.15  &  2  &  24.3  &  \ldots & 24.3 & \ldots &\ldots & \ldots &         1.81  &    \ldots   \\
\enddata
\tablecomments{Table~\ref{tbl:catalog} is published in its entirety in the electronic version of the {\em Astrophysical Journal}. 
A portion is shown here for guidance regarding its form and content.}
\tablenotetext{a} {X-ray positions are corrected for average X-ray to optical offset (to correct {\em Chandra} astrometry).}
\tablenotetext{b} {Note that the $HR$s presented here for sources with soft-band counts that did not meet our detection criteria are calculated differently than they were in Paper I. Please see \S\ref{sec:catalog} for detail.}
\tablenotetext{c} {See \S\ref{sec:catalog} for details of the flag code. Briefly: 0 = no optical coverage, 
1 = solid optical ID, 
2 = upper limit, 
3 = saturated in SEXSI optical image; $R$-band magnitude taken from the Guide Star Catalog II \citep{McLean:00},
4 = solid ID but $R$-band magnitude affected by nearby bright source, 
5 = more than one optical source consistent with the X-ray source position, 
6 = upper limit (no optical counterpart) but area contaminated by nearby bright source so limiting magnitude is unknown.}
\tablenotetext{d} {$HR$ is lower limit.}
\tablenotetext{e} {Source falls within an area potentially less than 1~Mpc from a target cluster center. Source was not used for \hardrange~\lognlogs~calculation in Paper 1.}
\label{tbl:catalog}
\end{deluxetable}

\clearpage

\end{document}